\documentclass[9pt,twocolumn,twoside]{opticajnl}
\journal{opticajournal} 
\setboolean{shortarticle}{false}
\usepackage{lineno}
\usepackage{graphicx}
\usepackage{dcolumn}
\usepackage{amsmath,amsfonts,amssymb}
\usepackage{bm}
\usepackage{lipsum}
\usepackage[colorlinks=true, allcolors=blue]{hyperref}
\usepackage{url}
\usepackage{lineno}
\usepackage{booktabs}
\usepackage{tabularx}
\usepackage{amsmath}

\title{Phase-enhanced nonreciprocal photon-phonon conversion via coupled optomechanical cavities}

\author{Divya Mishra}
\author[1,*]{Parvendra Kumar}

\affil{Optics  and  Photonics Centre, Indian Institute of Technology Delhi, Hauz Khas, New Delhi-110016, India}

\affil[*]{parvendra@iitd.ac.in}
\setboolean{displaycopyright}{false}

\begin{abstract}
Nonreciprocity, characterized by direction-dependent signal propagation, is fundamental to technologies such as isolators, signal routing, and precision sensing. This letter theoretically demonstrates nonreciprocal phonon transport and the conversion between photon and acoustic phonon signals in coupled optomechanical cavities via phase-dependent driving. It is demonstrated that, in contrast to nonreciprocal phonon transport, which necessitates both dissipation and phase-induced violation of time reversal symmetry, the nonreciprocity in photon-phonon conversion can occur without violating time reversal symmetry. We demonstrate that such nonreciprocity arises due to the path-dependent asymmetry in photon-phonon conversion. Furthermore, we demonstrate that the nonreciprocity of photon-phonon conversion can be further enhanced, achieving isolation levels of up to 40 dB by suitably modifying the phase difference of the driving lasers.
\end{abstract}

\begin{document}

\maketitle

\section{Introduction}\label{sec:level1}
Controlling the directional transmission of photonic and phononic signals in integrated platforms is of fundamental and practical importance in contemporary photonics and cavity optomechanics~\cite{ref1, ref2, ref3, ref4,ref5}. Non-reciprocal devices, such as isolators and circulators, are essential for signal processing and quantum technologies~\cite{ref6, ref7}. Traditional approaches rely on magneto-optical materials that disrupt time-reversal symmetry through external magnetic fields. However, these methods face limitations due to the weak response of such materials at optical frequencies and their limited compatibility with integrated platforms. This has prompted the investigation of alternative mechanisms that enable scalable, on-chip non-reciprocity. Such mechanisms achieve non-reciprocity through engineered interference and dynamic modulation, allowing for the creation of effective gauge fields without the need for magnetic materials. In cavity optomechanical devices, synthetic gauge fields provide a natural platform for the non-reciprocal transport of photons and phonons, as coherent photon-phonon interactions enable tunable mode hybridization and energy transfer~\cite{ref8, ref9, ref10, ref11}. Recent experimental demonstrations show that in coupled optomechanical cavities, phase-controlled optical drives can induce complex hopping amplitudes across sites. This results in synthetic magnetic flux, leading to the violation of time-reversal symmetry (TRS). The violation of TRS, combined with the dissipation of mechanical modes, facilitates the directional transport of photons, offering a pathway for reconfigurable non-reciprocal functionality in fully integrated structures~\cite{ref12, ref13}. Additionally, the conversion between photonic and phononic signals is crucial for transduction-based devices and applications, such as mass sensing ~\cite{ref14}. Recently, the quantum transduction of microwave photons to optical photons and qubit read out have also been demonstrated using hybrid optomechanical systems~\cite{ref15, ref16, ref17}.

\vspace{0.2cm}

In this letter, we theoretically investigate non-reciprocal phonon transport and the conversion between photon and phonon signals in coupled optomechanical cavities, which involve coupling between two optical modes and two mechanical modes. We demonstrate that nonreciprocity in phonon transport necessitates both phase-induced breaking of time-reversal symmetry at non-integer synthetic flux values and non-zero dissipation of the optical modes. However, we find that, in the case of photon-phonon conversion, breaking time-reversal symmetry via synthetic magnetic flux is not essential. Nonreciprocal photon-phonon conversion arises from interference between nonidentical paths during forward and backward conversion. Furthermore, we show that the phonon isolation—defined as the ratio of forward and backward transmission of phonons—can be significantly enhanced by appropriately choosing the value of synthetic flux. We reveal that phonon isolation and photon-phonon conversion isolation can be tuned by controlling synthetic flux through the phases of the driving lasers. With optimized synthetic flux and coupling rates between the optical and mechanical modes, we can achieve phonon isolation of 60 dB and photon-phonon conversion isolation of 40 dB.

\section{Theory}
\begin{figure}[htbp]
    \centering
    \includegraphics[width=0.4\textwidth]{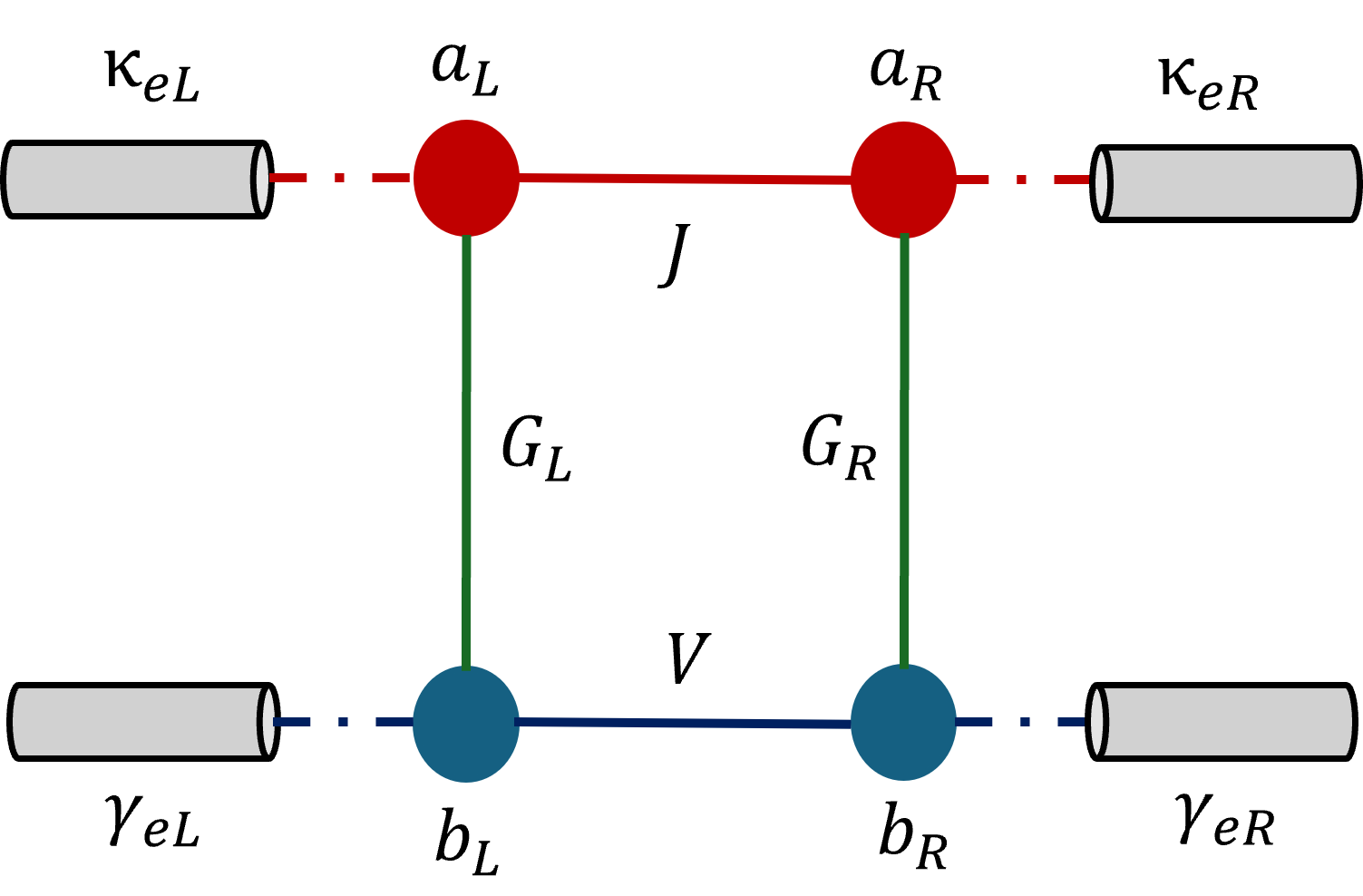}
    \caption{Schematic diagram of the two coupled optomechanical cavities. 
    The optical modes (red) are coupled via photon hopping $J$, 
    while the mechanical modes (blue) are coupled via phonon hopping $V$,  and the optomechanical coupling by $G_j$. The operators ${a}_j$ and ${b}_j$ represent the optical and mechanical field modes, respectively, while $\kappa_{e j}, \gamma_{e j}$  external optical and mechanical decay rates, respectively.}
    \label{fig:1}
\end{figure}

{\label{sec2}}
We analyze a system composed of two coupled cavities, referred to as the left $(L)$ and right $(R)$ cavities. Each cavity supports a localized optical mode and a mechanical mode, which are coupled through vacuum optomechanical interaction. Additionally, the system incorporates coupling between the optical and mechanical modes. This coupling can be achieved by physically connecting the two optomechanical cavities using waveguides~\cite{ref12}. Furthermore, the optical and mechanical modes are coupled to external waveguides to explore the coupling and transmission of photon and phonon signals ~\cite{ref9, ref12}. A schematic representation of the system is provided in Fig.~\ref{fig:1}.
The linearized Hamiltonian that describes the system, including the coupling with input photon and phonon signals, is given as
[Appendix~\ref{A}]
\begin{equation}
\begin{aligned}
H_{\mathrm{lin}} =\;
& -\sum_{j=L,R} \Delta_j\, a_j^\dagger a_j
+ J \left(a_L^\dagger a_R + a_R^\dagger a_L \right) \\
& + \sum_{j=L,R} \omega_{mj}\, b_j^\dagger b_j
+ V \left(b_L^\dagger b_R + b_R^\dagger b_L \right) \\
& + \sum_{j=L,R} G_j \left(
e^{i\phi_j} a_j b_j^\dagger
+ e^{-i\phi_j} a_j^\dagger b_j
\right) \\
& + i \sum_{j=L,R} \sqrt{\kappa_{ej}}\, \eta_j
\left(
a_j e^{i(\omega-\omega_d)t}
- a_j^\dagger e^{-i(\omega-\omega_d)t}
\right) \\
& + i \sum_{j=L,R} \sqrt{\gamma_{ej}}\, \zeta_j
\left(
b_j e^{i\omega t}
- b_j^\dagger e^{-i\omega t}
\right),
\end{aligned}
\label{eq:1}
\end{equation}
here, $\Delta_j = \omega_{dj} - \omega_{cj}=-\omega_{mj}$ denotes the detuning of the optical mode with respect to the driving laser frequency, where $\omega_{dj}$ is the frequency of the laser driving cavity $j \in \{L,R\}$ and $\omega_{cj}$ is the corresponding cavity resonance frequency. The mechanical resonance frequencies of the left and right cavities are given by $\omega_{mL}$ and $\omega_{mR}$, respectively. 

The operators $a_j$ ($a_j^\dagger$) and $b_j$ ($b_j^\dagger$) represent the annihilation (creation) operators for the optical (photon) and mechanical (phonon) modes. The parameter $J$ characterizes the coherent optical coupling between the two cavities, while $V$ denotes the coupling strength between the mechanical modes. The linearized optomechanical interaction is described by the enhanced coupling strength $G_j = g_j \alpha_j$, where $g_j$ is the vacuum optomechanical coupling rate and $\alpha_j$ is the steady-state intracavity field amplitude. Moreover, \( \eta_j \) and \( \zeta_j \) denote the input photon and phonon signal amplitudes coupled to the system via the external coupling rates \( \kappa_{ej} \) and \( \gamma_{ej} \), respectively.
\vspace{0.2cm}

\section{Results and Discussion}{\label{sec3}}

We investigate nonreciprocity in phonon transport and photon-phonon conversion by simulating the analytical results. In the simulation, we employed experimentally feasible parameters for the optomechanical system under consideration, which are detailed in Table ~\ref{table:1}~\cite{ref12}. 

\begin{table}[htbp]
\centering
\caption{Simulation parameters of the optomechanical system}
\label{table:1}
\renewcommand{\arraystretch}{1.0}

\begin{tabularx}{\columnwidth}{X c c}
\toprule
\textbf{Description} & \textbf{Parameter} & \textbf{Value} \\
\midrule
Coupling strength beween left and right optical modes & $J/2\pi$ & $110~\mathrm{MHz}$ \\
Optomechanical coupling strength & $G_{L(R)}$ & $33~(31)~\mathrm{MHz}$ \\
Mechanical mode frequencies & $\omega_{mL(mR)}/2\pi$ & $5.7884~(5.7791)~\mathrm{GHz}$ \\
External coupling rates of optical modes & $\kappa_{eL(R)}/2\pi$ & $0.74~(0.44)~\mathrm{GHz}$ \\
External coupling rates of mechanical modes & $\gamma_{eL(R)}/2\pi$ & $4.3~(5.7)~\mathrm{MHz}$ \\
Internal decay rates of optical modes& $\kappa_{L(R)}/2\pi$ & $0.29~(0.31)~\mathrm{GHz}$ \\
Internal decay rates of mechanical modes & $\gamma_{iL(R)}/2\pi$ & $1.0~(1.2)~\mathrm{MHz}$ \\
\bottomrule
\end{tabularx}

\end{table}

\subsection{Nonreciprocal transport of a phonon signal}{\label{sec:A}}
The nonreciprocal transport of phonons is quantified by the isolation, defined as the ratio of forward to backward phonon transmission, and is given by Eq.~\eqref{eq:2} [Appendix~\ref{B}]. It is important to note that the optically mediated coupling $\Gamma_{A}$ between the two mechanical modes becomes purely real when $\kappa_R = \kappa_L = 0$ for all frequencies. In this limit, the system remains reciprocal despite the breaking of time-reversal symmetry via the synthetic flux $\phi$, since $\left| V - \Gamma_{A} e^{-i\phi} \right| = \left| V - \Gamma_{A} e^{i\phi} \right|$. Therefore, achieving nonreciprocity in phonon transport requires both nonzero optical decay rates and the breaking of time-reversal symmetry.
\begin{equation}
I_{\text{phonon}} =
10 \log_{10}\Bigg[\frac{
\left| V - {\Gamma_{A} e^{-i\phi}} \right|^2
}{
\left| V -  {\Gamma_{A} e^{i\phi}} \right|^2
}
\Bigg],
\label{eq:2}
\end{equation}
where,
$\Gamma_{A}$$=$$\dfrac{J G_L G_R}{\Delta_A}$, $\Delta_A$$=$$\chi_{aR}^{-1}\chi_{aL}^{-1} + J^2$, $\chi_{aL}^{-1}$$=$$-i(\omega + \Delta_L) + \frac{\kappa_L}{2}$,
$\chi_{aR}^{-1}$$ = $$ -i(\omega + \Delta_R) + \frac{\kappa_R}{2}$ and $\phi$$ = $$\phi_L-\phi_R$ is the synthetic flux threading the four-mode plaquette shown in Fig.~\ref{fig:1}.
\begin{figure}[!t]
    \centering
    \includegraphics[width=0.4\textwidth]{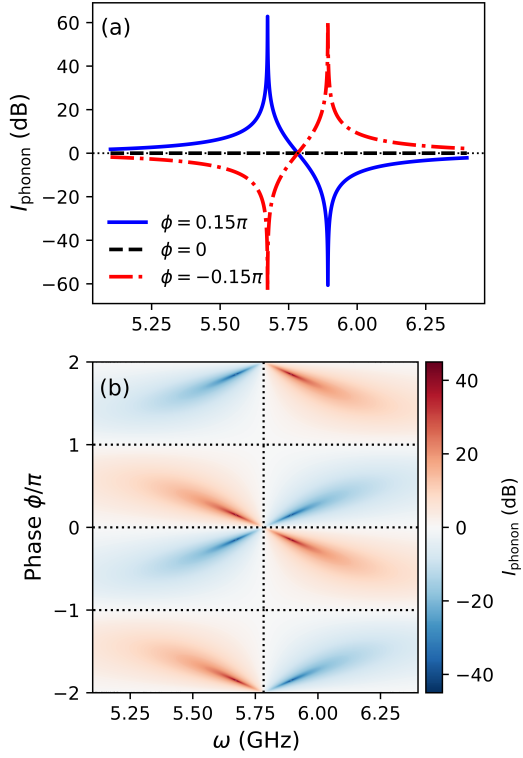}
    \caption{(a) Isolation describing the ratio of forward and backward phonon signal transport for three different values of the synthetic flux $\phi$, and (b) Dependence of the phonon isolation on the synthetic flux $\phi$ and frequency $\omega$.}
    \label{fig:2}
\end{figure}

To maximize the phonon isolation, we optimize the magnitude and phase parts of $\Gamma_{A}$ such that nearly perfect destructive interference occurs in $\left| V - \Gamma_{A} e^{i\phi} \right|$, yielding a maximum isolation of approximately 60~dB, as shown in Fig.~\ref{fig:2}(a). It is also evident from Figs.~\ref{fig:2}(a) and \ref{fig:2}(b) that for synthetic flux values $\phi = 0$ or $\phi = n\pi$, the isolation reduces to 0~dB, rendering the system reciprocal due to the preservation of time-reversal symmetry. Furthermore, the isolation is fully controllable via the synthetic flux $\phi$, allowing the system to be switched between reciprocal and nonreciprocal regimes solely by tuning the phase of the driving lasers. In Fig.~\ref{fig:2}(b), the isolation of phonon transport is plotted as a function of the signal frequency $\omega$ and the synthetic flux $\phi$, further demonstrating the tunability of nonreciprocity through phase control.
\vspace{0.2cm}
\subsection{Nonreciprocal photon-to-phonon conversion}
\begin{figure}[t]
    \centering
    \includegraphics[width=0.4\textwidth]{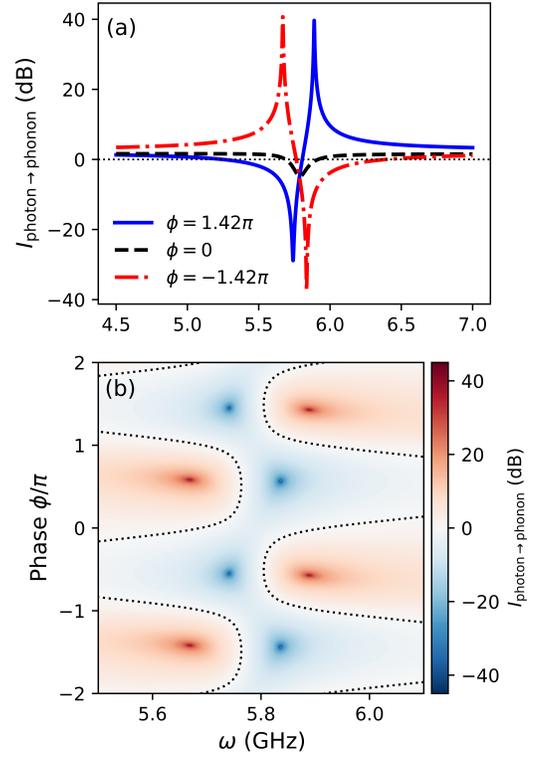}
    \caption{(a) Isolation describing the ratio of the forward and backward nonreciprocal conversion of an input photon signal into a phonon signal for three different values of the synthetic flux $\phi$. (b) Dependence of the photon-to-phonon conversion on the synthetic flux 
$\phi$ and frequency $\omega$, demonstrating phase-controlled nonreciprocity.}
    \label{fig:3}
\end{figure}

{\label{sec:B}}
The ratio of forward ($L \rightarrow R$) and backward ($R \rightarrow L$) photon-to-phonon conversion defines the isolation, which quantifies the degree of nonreciprocity in the conversion process and is given by [Appendix~\ref{B}]
\begin{equation}
\begin{aligned}
I_{\mathrm{photon\rightarrow phonon}}
=
10 \log_{10}\!\Bigg[
\frac{
|\chi_{a_R}^{-1}|^2 |G_L|^2
\left|
V + \Gamma_- e^{-i\phi}
\right|^2
}{
|\chi_{a_L}^{-1}|^2 |G_R|^2
\left|
V + \Gamma_+ e^{i\phi}
\right|^2
}
\Bigg],
\end{aligned}
\label{eq:3}
\end{equation}
where, $\Gamma_-$$=$$\frac{J}{\chi_{a_R}^{-1}}\frac{G_R}{G_L}\chi_{b_L}^{-1}$, $\Gamma_+$$=$$\frac{J}{\chi_{a_L}^{-1}}\frac{G_L}{G_R}\chi_{b_R}^{-1}$, $\chi_{aL}^{-1}$$=$$-i(\omega + \Delta_L) + \frac{\kappa_L}{2}$, $\chi_{aR}^{-1}$$ =$$ -i(\omega + \Delta_R) + \frac{\kappa_R}{2}$ and $\chi_{bL}^{-1}$$ = $$-i(\omega - \omega_{mL}) + \frac{\gamma_L}{2} $, $\chi_{bR}^{-1} $$= $$-i(\omega - \omega_{mR}) + \frac{\gamma_R}{2}$.
A careful inspection of Eq.~\eqref{eq:3} reveals that the forward conversion process, proportional to $|\chi_{a_R}^{-1}|^2 |G_L|^2 \left| V + \Gamma_- e^{-i\phi} \right|$, differs from the backward conversion process, proportional to $|\chi_{a_L}^{-1}|^2 |G_R|^2 \left| V + \Gamma_+ e^{i\phi} \right|$. This asymmetry in the conversion pathways gives rise to nonreciprocity, as evidenced by the nonzero isolation in Figs.~\ref{fig:3}(a) and (b), even for $\phi = 0$. To maximize the isolation, we optimize the magnitude and phase parts of $\Gamma_{+}$ such that the term $\left| V + \Gamma_+ e^{i\phi} \right|$ undergoes destructive interference at a specific frequency. As shown in Fig.~\ref{fig:3}(a), the isolation reaches a maximum value of approximately $40~\mathrm{dB}$ around $5.9~\mathrm{GHz}$ for $\phi = 1.42\pi$. For the opposite synthetic flux, $\phi = -1.42\pi$, the isolation becomes negative, indicating enhanced backward photon-to-phonon conversion and demonstrating the tunability of isolation via phase control. Figure~\ref{fig:3}(b) shows the isolation of photon-to-phonon conversion as a function of frequency $\omega$ and synthetic flux $\phi$. In contrast to phonon isolation, nonreciprocal photon-to-phonon conversion can be achieved even without breaking time-reversal symmetry, i.e., for $\phi = 0$ or $\phi = n\pi$, although with reduced isolation.

\vspace{0.3cm}
\subsection{Nonreciprocal phonon-to-photon conversion}{\label{sec:C}}
The isolation, which quantifies the degree of nonreciprocity in the phonon-to-photon conversion process, is defined as the ratio of forward ($L \rightarrow R$) and backward ($R \rightarrow L$) phonon-to-photon conversion and is given by [Appendix~\ref{B}]
\begin{equation}
\begin{aligned}
I_{\mathrm{phonon\rightarrow photon}}
=
10 \log \Bigg[
\frac{|\chi_{a_L}^{-1}|^2 |G_R|^2
\left|V+\Gamma_+e^{-i\phi}\right|^2}{
|\chi_{a_R}^{-1}|^2 |G_L|^2\left|V+\Gamma_-e^{i\phi}\right|^2}
\Bigg],
\end{aligned}
\end{equation}
where, $\Gamma_-$$=$$\frac{J}{\chi_{a_R}^{-1}}\frac{G_R}{G_L}\chi_{b_L}^{-1}$,$\Gamma_+$$=$$\frac{J}{\chi_{a_L}^{-1}}\frac{G_L}{G_R}\chi_{b_R}^{-1}$,$\chi_{aL}^{-1}$$=$$-i(\omega+\Delta_L)+\frac{\kappa_L}{2}$,$\chi_{aR}^{-1}$$=$$ -i(\omega+\Delta_R) + \frac{\kappa_R}{2}$and $\chi_{bL}^{-1}$$=$$-i(\omega -\omega_{mL})+\frac{\gamma_L}{2} $,$\chi_{bR}^{-1}$$=$$-i(\omega - \omega_{mR})+\frac{\gamma_R}{2}$. 
\vspace{0.1cm}
\begin{figure}[htbp]
    \centering
    \includegraphics[width=0.4\textwidth]{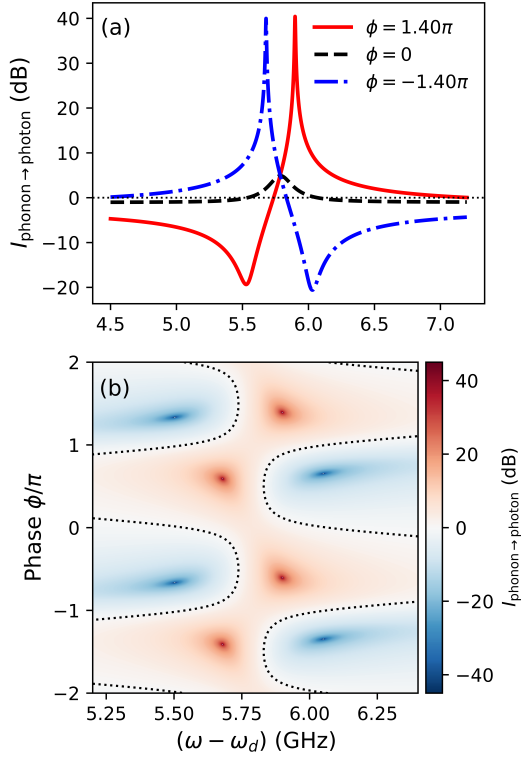}
    \caption{(a) Isolation describing the ratio of the forward and backward nonreciprocal conversion of an input phonon signal into an optical photon signal for three different values of the synthetic flux $\phi$. (b) Dependence of the phonon-to-photon conversion on the synthetic flux 
$\phi$ and frequency $(\omega-\omega_{d})$, demonstrating phase-controlled nonreciprocity.}
    \label{fig:4}
\end{figure}

It is evident from Fig.~\ref{fig:4}(a) that positive isolation, indicating nonreciprocity in phonon-to-photon conversion, can be achieved even for $\phi = 0$. This arises solely from the asymmetry between the forward $\left(b_{L} \rightarrow b_{R} \rightarrow a_{R} + b_{L} \rightarrow a_{L} \rightarrow a_{R}\right)$ and backward $\left(b_{R} \rightarrow b_{L} \rightarrow a_{L} + b_{R} \rightarrow a_{R} \rightarrow a_{L}\right)$ conversion pathways. Similar to the photon-to-phonon case, we optimize the real and imaginary parts of $\Gamma_{-}$ such that the term $\left| V + \Gamma_- e^{i\phi} \right|$ undergoes destructive interference at a specific frequency. For $\phi = 1.4\pi$, the maximum positive isolation occurs near $(\omega - \omega_{d}) \approx 6~\mathrm{GHz}$. However, for $\phi = -1.4\pi$, the isolation becomes negative, as shown in Fig.~\ref{fig:4}(a), indicating enhanced backward conversion and demonstrating control over isolation via tuning the relative phase of the driving lasers. Figure~\ref{fig:4}(b) presents the isolation of phonon-to-photon conversion as a function of frequency $(\omega - \omega_{d})$ and synthetic flux $\phi$, further confirming the tunability of nonreciprocity through phase control.

\section {Conclusion}{\label{sec4}}
In conclusion, we have theoretically investigated nonreciprocal phonon transport as well as photon-to-phonon and phonon-to-photon conversion in coupled optomechanical cavities. We show that achieving nonreciprocity in phonon transport requires both nonzero optical decay rates and the breaking of time-reversal symmetry. In contrast, the nonreciprocity in photon-to-phonon and phonon-to-photon conversion can be achieved even without breaking time-reversal symmetry, originating from the intrinsic asymmetry between forward and backward conversion pathways. Furthermore, the degree of nonreciprocity can be significantly enhanced by optimizing the synthetic flux and coupling parameters. Under optimal conditions, we obtain phonon isolation up to $60~\mathrm{dB}$ and photon--phonon conversion isolation up to $40~\mathrm{dB}$. These results can be useful for realizing on-chip, phase-programmable nonreciprocal devices, with potential applications in quantum information processing and integrated phononic and photonic circuits.

\textbf{Acknowledgements}

\indent D.~M. acknowledges the financial support in the form of a research fellowship from the Indian Institute of Technology Delhi, New Delhi. P.~K. acknowledges the support through the Seed Grant for New Faculty (MI02921) from the Indian Institute of Technology Delhi. The authors also thank Amarendra Kumar Sarma for insightful discussions.

\appendix
\renewcommand{\theequation}{A\arabic{equation}}
\setcounter{equation}{0}
\section{APPENDIX A: DERIVATION OF THE LINEARIZED HAMILTONIAN}{\label{A}}
To derive the linearized Hamiltonian given in the main text ~\eqref{eq:1}, we begin with the Hamiltonian describing two coupled optomechanical cavities, including both optomechanical interactions and inter-cavity coupling between the optical and mechanical modes. We consider that both optical modes are driven by phase-tunable continuous-wave laser fields. In this setting, the Hamiltonian reads as
\begin{equation}
\begin{aligned}
H =&\ \hbar \sum_{j=L,R} \omega_{cj}\, a_j^\dagger a_j
+ \hbar J \left(a_L^\dagger a_R + a_R^\dagger a_L \right)
+ \hbar \sum_{j=L,R} \omega_{mj}\, b_j^\dagger b_j \\
&+ \hbar V \left(b_L^\dagger b_R + b_R^\dagger b_L \right)
+ \hbar \sum_{j=L,R} g_j\, a_j^\dagger a_j \left(b_j + b_j^\dagger\right) \\
&+ i\hbar \sum_{j=L,R} \sqrt{\kappa_{ej}}\, \epsilon_j
\left(
a_j^\dagger e^{-i(\omega_{dj} t+\phi_j)}
- a_j e^{i(\omega_{dj} t+\phi_j)}
\right),
\end{aligned}
\end{equation}
here, \( \omega_{cL} \) and \( \omega_{cR} \) are the optical resonance frequencies of the left and right cavities, and \( \omega_{mL} \), \( \omega_{mR} \) are the corresponding mechanical mode frequencies. The operators \( a_j \) (\( a_j^\dagger \)) and \( b_j \) (\( b_j^\dagger \)) denote the annihilation (creation) operators for photons and phonons, respectively. The parameter \( J \) describes the optical coupling (photon tunneling) between the two cavities, while \( V \) characterizes the coupling between the mechanical modes (phonon tunneling), and parameter \( \epsilon_{j} \) denoting laser amplitude coupled to the optical modes via the external rate \( \kappa_{ej} \). Next, we transform the Hamiltonian into a frame rotating with the frequency of driving laser as 
$H_{\mathrm{RF}} = U H U^{\dagger} - i\hbar\, U \frac{dU^{\dagger}}{dt}$, 
where $U$ is the unitary transformation operator. The unitary operator is defined as 
$U = \exp\!\left(i \sum_{j=L,R} \omega_{d} t\, a_j^{\dagger} a_j \right)$, 
which corresponds to a frame rotating at the driving  frequency $\omega_{d}$ of the cavity modes.
\begin{equation}
\begin{aligned}
H_{\mathrm{RF}} =\;
& - \hbar \sum_{j=L,R} \Delta_j\, a_j^{\dagger} a_j
+ J \left(a_L^{\dagger} a_R + a_R^{\dagger} a_L \right) \\
& + \hbar \sum_{j=L,R} \omega_{mj}\, b_j^{\dagger} b_j
+ V \left(b_L^{\dagger} b_R + b_R^{\dagger} b_L \right) \\
& + \hbar \sum_{j=L,R} g_j\, a_j^{\dagger} a_j \left(b_j + b_j^{\dagger} \right) \\
& + i \hbar \sum_{j=L,R} \sqrt{\kappa_{ej}}\, \epsilon_j
\left(a_j e^{i\phi_j} - a_j^{\dagger} e^{-i\phi_j} \right),
\end{aligned}
\end{equation}
where, $\Delta_j$ is the detuning between the laser drive and cavity modes. The detuning parameter is defined as $\Delta_j$ = $\omega_{d}-\omega_{cj}$. The dynamics of the system can be observed through the Quantum Langevin Equations(QLEs). To derive the QLEs we are using the Heisenberg equation of motions,$\frac{dO}{dt} =-i [O , H_{\mathrm{RF}}],O\in a_j, b_j $ by including the input noises and decay rates phenomenally:
\begin{align}
\dot{a}_L &= \left(i\Delta_L-\frac{\kappa_L}{2}\right)a_L - iJ a_R
- i g_L a_L\left(b_L+b_L^{\dagger}\right) \nonumber\\
&\quad + \sqrt{\kappa_{eL}}\,\varepsilon_L e^{i\phi_L}
+ \sqrt{\kappa_{eL}}\,a_{L,\mathrm{in}}
+ \sqrt{\kappa_{iL}}\,a_{L,\mathrm{in}}^{(i)},
\\[4pt]
\dot{a}_R &= \left(i\Delta_R-\frac{\kappa_R}{2}\right)a_R - iJ a_L
- i g_R a_R\left(b_R+b_R^{\dagger}\right) \nonumber\\
&\quad + \sqrt{\kappa_{eR}}\,\varepsilon_R e^{i\phi_R}
+ \sqrt{\kappa_{eR}}\,a_{R,\mathrm{in}}
+ \sqrt{\kappa_{iR}}\,a_{R,\mathrm{in}}^{(i)},
\\[4pt]
\dot{b}_L &= -\left(i\omega_{mL}+\frac{\gamma_L}{2}\right)b_L - iV b_R
- i g_L a_L^{\dagger}a_L \nonumber\\
&\quad + \sqrt{\gamma_{eL}}\,b_{L,\mathrm{in}}
+ \sqrt{\gamma_{iL}}\,b_{L,\mathrm{in}}^{(i)},
\\[4pt]
\dot{b}_R &= -\left(i\omega_{mR}+\frac{\gamma_R}{2}\right)b_R - iV b_L
- i g_R a_R^{\dagger}a_R \nonumber\\
&\quad + \sqrt{\gamma_{eR}}\,b_{R,\mathrm{in}}
+ \sqrt{\gamma_{iR}}\,b_{R,\mathrm{in}}^{(i)},
\end{align}
where $\kappa_j=\kappa_{ej}+\kappa_{ij}$ and $\gamma_j=\gamma_{ej}+\gamma_{ij}$ for $j=L,R$. The external decay rates $\kappa_{e j}$ and $\gamma_{e j}$ define the input-output channels, while the internal rates $\kappa_{i j}$ and $\gamma_{i j}$ account for intrinsic loss and contribute only to dissipation and noise. The operators $a_{j,\mathrm{in}}$ and $b_{j,\mathrm{in}}$ denote the external input noise associated with the coupling rates $\kappa_{ej}$ and $\gamma_{ej}$, respectively, while $a_{j,\mathrm{in}}^{(i)}$ and $b_{j,\mathrm{in}}^{(i)}$ describe the internal noise arising from intrinsic losses $\kappa_{ij}$ and $\gamma_{ij}$. To obtain the linearized QLEs , we expand the operators as $a_j=\alpha_je^{-i\phi_j}+\delta a_j$ and $b_j=\beta_j+\delta b_j$ around their steady-state mean values $\alpha_je^{-i\phi_j}$ and $\beta_j$, while $\delta a_j$ and $\delta b_j$ are small quantum fluctuation around theses steady-state values.
\begin{align}
\alpha_L &=
\frac{
\left(\frac{\kappa_R}{2}-i\Delta_R'\right)\sqrt{\kappa_{eL}}\,\varepsilon_L e^{2i\phi_L}
- iJ\,\sqrt{\kappa_{eR}}\,\varepsilon_R e^{i(\phi_L+\phi_R)}
}{
\left(\frac{\kappa_L}{2}-i\Delta_L'\right)\left(\frac{\kappa_R}{2}-i\Delta_R'\right)+J^2
},
\\[6pt]
\alpha_R &=
\frac{
\left(\frac{\kappa_L}{2}-i\Delta_L'\right)\sqrt{\kappa_{eR}}\,\varepsilon_R e^{2i\phi_R}
- iJ\,\sqrt{\kappa_{eL}}\,\varepsilon_L e^{i(\phi_L+\phi_R)}
}{
\left(\frac{\kappa_L}{2}-i\Delta_L'\right)\left(\frac{\kappa_R}{2}-i\Delta_R'\right)+J^2
},
\end{align}
where, $\Delta_L' = \Delta_L - g_L \left(\beta_L + \beta_L^{*}\right)$ and 
\(
\Delta_R' = \Delta_R - g_R \left(\beta_R + \beta_R^{*}\right).
\)
The effective detuning \( \Delta'_j \) can be approximated as \( \Delta_j \), since the vacuum optomechanical coupling strength \( g_j \) is negligibly small.
We drive the linearized Hamiltonian by following ~\cite{ref1, ref18} and including the input photon and photon signals. For red-detuned laser-cavity coupling $\Delta_j$ = -$\omega_{mj}$, it is given as:
\begin{equation}
\begin{aligned}
H_{\mathrm{lin}} =\;
& -\sum_{j=L,R} \Delta_j\, a_j^\dagger a_j
+ J \left(a_L^\dagger a_R + a_R^\dagger a_L \right) \\
& + \sum_{j=L,R} \omega_{mj}\, b_j^\dagger b_j
+ V \left(b_L^\dagger b_R + b_R^\dagger b_L \right) \\
& + \sum_{j=L,R} G_j \left(
e^{i\phi_j} a_j b_j^\dagger
+ e^{-i\phi_j} a_j^\dagger b_j
\right) \\
& + i \sum_{j=L,R} \sqrt{\kappa_{ej}}\, \eta_j
\left(
a_j e^{i(\omega-\omega_d)t}
- a_j^\dagger e^{-i(\omega-\omega_d)t}
\right) \\
& + i \sum_{j=L,R} \sqrt{\gamma_{ej}}\, \zeta_j
\left(
b_j e^{i\omega t}
- b_j^\dagger e^{-i\omega t}
\right),
\end{aligned}
\label{eq:A9}
\end{equation}
here, \( \eta_j \) and \( \zeta_j \) denote the input photon and phonon signal amplitudes coupled to the system via the external rates \( \kappa_{ej} \) and \( \gamma_{ej} \), respectively. The linearized optomechanical coupling strength is \( G_j=\alpha_j g_j \), where \( g_j \) is the vacuum coupling rate and \( \alpha_j \) is the steady-state intracavity field amplitude.

\section{APPENDIX B: Nonreciprocal Transport and Conversion}{\label{B}}
\renewcommand{\theequation}{B\arabic{equation}}
\setcounter{equation}{0}
 To analyze the non-reciprocal transmission or photon-phonon conversion, we drive the quantum Langevin equations (QLEs) without including the input noises and then transform them to the frequency domain. The QLEs in frequency domain can be written in matrix form as $M(\omega)\, O(\omega) = \Omega N_{in}$. These matrices are defined as
    The frequency-domain equations can be written as 
\( M(\omega)\,O(\omega)=\Omega\,N_{\mathrm{in}}(\omega) \), 
where
\[
M(\omega)=
\begin{bmatrix}
\chi_{a_L}^{-1} & iJ & iG_L e^{-i\phi_L} & 0 \\
iJ & \chi_{a_R}^{-1} & 0 & iG_R e^{-i\phi_R} \\
iG_L e^{i\phi_L} & 0 & \chi_{b_L}^{-1} & iV \\
0 & iG_R e^{i\phi_R} & iV & \chi_{b_R}^{-1}
\end{bmatrix}.
\]
Here, \( O(\omega)=(a_L,a_R,b_L,b_R)^{\mathrm{T}} \), 
\( N_{\mathrm{in}}(\omega)=(\eta_L,\eta_R,\zeta_L,\zeta_R)^{\mathrm{T}} \), and 
\( \Omega=\mathrm{diag}(\sqrt{\kappa_{eL}},\sqrt{\kappa_{eR}},\sqrt{\gamma_{eL}},\sqrt{\gamma_{eR}}) \).

The matrix \(M(\omega)\) can be decomposed into \(2\times2\) blocks as 
\( M(\omega)=\begin{bmatrix} A & C \\ D & B \end{bmatrix} \), 
with
\( A=\begin{bmatrix}\chi_{a_L}^{-1} & iJ \\ iJ & \chi_{a_R}^{-1}\end{bmatrix} \), 
\( B=\begin{bmatrix}\chi_{b_L}^{-1} & iV \\ iV & \chi_{b_R}^{-1}\end{bmatrix} \), 
\( C=\begin{bmatrix} iG_L e^{-i\phi_L} & 0 \\ 0 & iG_R e^{-i\phi_R} \end{bmatrix} \), and 
\( D=\begin{bmatrix} iG_L e^{i\phi_L} & 0 \\ 0 & iG_R e^{i\phi_R} \end{bmatrix} \).
 where as $\chi_{aL}^{-1}$$ = -i(\omega + \Delta_L) + \frac{\kappa_L}{2}$ and $\chi_{aR}^{-1} $$= -i(\omega + \Delta_R) + \frac{\kappa_R}{2}$ , $\chi_{bL}^{-1} $$= -i(\omega -\omega_{mL}) + \frac{\gamma_L}{2}$ and $\chi_{bR}^{-1}$$= -i(\omega - \omega_{mR}) + \frac{\gamma_R}{2}$. Using the block inversion formula, the inverse of \(M(\omega)\) can be written as
\[
M^{-1}(\omega)=
\begin{bmatrix}
A_{\mathrm{eff}}^{-1} & -A_{\mathrm{eff}}^{-1} C B^{-1} \\
- B_{\mathrm{eff}}^{-1} D A^{-1} & B_{\mathrm{eff}}^{-1}
\end{bmatrix},
\] where \( A_{\mathrm{eff}}^{-1} = [A - C B^{-1} D]^{-1} \) and 
\( B_{\mathrm{eff}}^{-1} = [B - D A^{-1} C]^{-1} \).
\vspace{0.1cm}
The effective optical block reads
\[
A_{\mathrm{eff}}^{-1}
=
\frac{1}{\Delta_{A_{\mathrm{eff}}}}
\begin{bmatrix}
\chi_{a_R}^{-1} + \dfrac{G_R^2 \chi_{b_L}^{-1}}{\Delta_B} &
-i\!\left(J - \dfrac{V G_L G_R e^{-i\phi}}{\Delta_B}\right) \\
-i\!\left(J - \dfrac{V G_L G_R e^{i\phi}}{\Delta_B}\right) &
\chi_{a_L}^{-1} + \dfrac{G_L^2 \chi_{b_R}^{-1}}{\Delta_B}
\end{bmatrix},
\]
while the effective mechanical block is
\[
B_{\mathrm{eff}}^{-1}
=
\frac{1}{\Delta_{B_{\mathrm{eff}}}}
\begin{bmatrix}
\chi_{b_R}^{-1} + \dfrac{G_R^2 \chi_{a_L}^{-1}}{\Delta_A} &
-i\!\left(V - \dfrac{J G_L G_R e^{i\phi}}{\Delta_A}\right) \\
-i\!\left(V - \dfrac{J G_L G_R e^{-i\phi}}{\Delta_A}\right) &
\chi_{b_L}^{-1} + \dfrac{G_L^2 \chi_{a_R}^{-1}}{\Delta_A}
\end{bmatrix}.
\]
Where, $\Delta_{Aeff}$ $=$ $(\chi_{a_L}^{-1}+\dfrac{G_L^2 \chi_{b_R}^{-1}}{\Delta_B})(\chi_{a_R}^{-1}+\dfrac{G_R^2 \chi_{b_L}^{-1}}{\Delta_B})+(J-\dfrac{V G_L G_R e^{i\phi}}{\Delta_B})(J-\dfrac{V G_L G_R e^{-i\phi}}{\Delta_B})$ and $\Delta_B$$ = $$(\chi_{bR}^{-1}\chi_{bL}^{-1}+V^2)$ , $\Delta_{Beff}$ $ = $ $(\chi_{b_L}^{-1}+\dfrac{G_L^2 \chi_{a_R}^{-1}}{\Delta_A})(\chi_{b_R}^{-1}+\dfrac{G_R^2 \chi_{a_L}^{-1}}{\Delta_A})+(V-\dfrac{J G_L G_R e^{i\phi}}{\Delta_A})(V-\dfrac{J G_L G_R e^{-i\phi}}{\Delta_A})$, and $\Delta_A$$ = $$(\chi_{aR}^{-1}\chi_{aL}^{-1}+J^2)$ . The phonon isolation, defined as the ratio of forward and backward phonon transport, is determined by the ratio of the off-diagonal elements of the effective mechanical block, \( B_{\mathrm{eff}}^{-1}\big|_{21} / B_{\mathrm{eff}}^{-1}\big|_{12} \). It reads as
\begin{equation}
I_{\mathrm{phonon}} =
10 \log \left[\frac{\left| V - \dfrac{J G_L G_R e^{-i\phi}}{\Delta_A} \right|^2}
{\left| V - \dfrac{J G_L G_R e^{i\phi}}{\Delta_A} \right|^2}
\right]
\end{equation}
Similarly, the isolation associated with photon-to-phonon and phonon-to-photon conversion is determined by the ratios \( \big(-A_{\mathrm{eff}}^{-1} C B^{-1}\big)\big|_{21} / \big(-A_{\mathrm{eff}}^{-1} C B^{-1}\big)\big|_{12} \) and \( \big(-B_{\mathrm{eff}}^{-1} D A^{-1}\big)\big|_{21} / \big(-B_{\mathrm{eff}}^{-1} D A^{-1}\big)\big|_{12} \), respectively. They read as

\begin{equation}
I_{\mathrm{photon}\rightarrow\mathrm{phonon}} =
10 \log \left[
\frac{
\left| \chi_{a_R}^{-1} \right|^2 |G_L|^2 
\left| V + \dfrac{J}{\chi_{a_R}^{-1}} \dfrac{G_R}{G_L} \chi_{b_L}^{-1} e^{-i\phi} \right|^2
}{
\left| \chi_{a_L}^{-1} \right|^2 |G_R|^2 
\left| V + \dfrac{J}{\chi_{a_L}^{-1}} \dfrac{G_L}{G_R} \chi_{b_R}^{-1} e^{i\phi} \right|^2
}
\right].
\end{equation}

\begin{equation}
I_{\mathrm{phonon}\rightarrow\mathrm{photon}} =
10 \log \left[
\frac{
\left| \chi_{a_L}^{-1} \right|^2 |G_R|^2 
\left| V + \dfrac{J}{\chi_{a_L}^{-1}} \dfrac{G_L}{G_R} \chi_{b_R}^{-1} e^{-i\phi} \right|^2
}{
\left| \chi_{a_R}^{-1} \right|^2 |G_L|^2 
\left| V + \dfrac{J}{\chi_{a_R}^{-1}} \dfrac{G_R}{G_L} \chi_{b_L}^{-1} e^{i\phi} \right|^2
}
\right].
\end{equation}


\begin{thebibliography}{99}


\bibitem{ref1}
M.~Aspelmeyer, T.~J.~Kippenberg, and F.~Marquardt, “Cavity optomechanics,” \href{https://doi.org/10.1103/RevModPhys.86.1391}{\textit{Rev. Mod. Phys.} \textbf{86}, 1391--1452 (2014).}

\bibitem{ref2}
A.~K.~Sarma and S.~Kalita, "Tutorial: Cavity Quantum Optomechanics,"
\href{https://doi.org/10.56042/ijpap.v61i7.103}{Indian J. Pure Appl. Phys. \textbf{61}(7), (2023).}

\bibitem{ref3}
N.~R.~Bernier, L.~D.~Tóth, A.~Koottandavida, M.~D.~Shah, M.~J.~H.~Kuzyk, S.~D.~Bennett, and K.~W.~Lehnert,
"Nonreciprocal reconfigurable microwave optomechanical circuit,"
\href{https://doi.org/10.1038/s41467-017-00447-1}{\textit{Nat. Commun.} \textbf{8}, 604 (2017).}

\bibitem{ref4}
A.~Seif, W.~DeGottardi, K.~Esfarjani, and M.~Hafezi,
"Thermal management and non-reciprocal control of phonon flow via optomechanics,"
\href{https://doi.org/10.1038/s41467-018-03624-y}{\textit{Nat. Commun.} \textbf{9}, 1207 (2018).}

\bibitem{ref5}
Z.~Shen, Y.~L.~Zhang, Y.~Chen, C.~L.~Zou, Y.~Guo, C.~H.~Dong, C.~W.~Qiu, and G.~C.~Guo,
"Reconfigurable optomechanical circulator and directional amplifier,"
\href{https://doi.org/10.1038/s41467-018-04187-8}{\textit{Nat. Commun.} \textbf{9}, 1797 (2018).}

\bibitem{ref6}
Z.~Shen, Y.~L.~Zhang, Y.~Chen, \textit{et al.}, “Experimental realization of optomechanically induced non-reciprocity,” \href{https://doi.org/10.1038/nphoton.2016.161}{\textit{Nat. Photonics} \textbf{10}, 657--661 (2016).}

\bibitem{ref7}
L.~Bi, J.~Hu, P.~Jiang, \textit{et al.}, “On-chip optical isolation in monolithically integrated non-reciprocal optical resonators,” 
\href{https://doi.org/10.1038/nphoton.2011.270}{\textit{Nat. Photonics} \textbf{5}, 758--762 (2011).}

\bibitem{ref8}
J.~P.~Mathew, J.~del~Pino, and E.~Verhagen, “Synthetic gauge fields for phonon transport in a nano-optomechanical system,” 
\href{https://doi.org/10.1038/s41565-019-0630-8}{\textit{Nat. Nanotechnol.} \textbf{15}, 198--202 (2020).}

\bibitem{ref9}
K.~Fang, M.~Matheny, X.~Luan, E.~H.~Kim, J.~Jiang, M.~J.~Burek, and P.~T.~Rakich,
"Optical transduction and routing of microwave phonons in cavity-optomechanical circuits,"
\href{https://doi.org/10.1038/nphoton.2016.107}{\textit{Nat. Photon.} \textbf{10}, 489--496 (2016).}

\bibitem{ref10}
Y.~Chen, Y.~L.~Zhang, Z.~Shen, \textit{et al.}, “Synthetic gauge fields in a single optomechanical resonator,” \href{https://doi.org/10.1103/PhysRevLett.126.123603}{\textit{Phys. Rev. Lett.} \textbf{126}, 123603 (2021).}

\bibitem{ref11}
Z.~Shen, Y.~L.~Zhang, Y.~Chen, Y.~F.~Xiao, C.~L.~Zou, G.~C.~Guo, and C.~H.~Dong,
"Nonreciprocal frequency conversion and mode routing in a microresonator,"
\href{https://doi.org/10.1103/PhysRevLett.130.013601}{\textit{Phys. Rev. Lett.} \textbf{130}, 013601 (2023).}

\bibitem{ref12}
K.~Fang, J.~Luo, A.~Metelmann, \textit{et al.}, “Generalized non-reciprocity in an optomechanical circuit via synthetic magnetism and reservoir engineering,” 
\href{https://doi.org/10.1038/nphys4009}{\textit{Nat. Phys.} \textbf{13}, 465--471 (2017).}

\bibitem{ref13}
A.~Metelmann and A.~A.~Clerk,
"Nonreciprocal photon transmission and amplification via reservoir engineering,"
\href{https://doi.org/10.1103/PhysRevX.5.021025}{\textit{Phys. Rev. X} \textbf{5}, 021025 (2015).}

\bibitem{ref14}
W.~Yu, W.~Jiang, Q.~Lin, and T.~Lu,
"Cavity optomechanical spring sensing of single molecules,"
\href{https://doi.org/10.1038/ncomms12311}{\textit{Nat. Commun.} \textbf{7}, 12311 (2016).}

\bibitem{ref15}
M.~Mirhosseini, A.~Sipahigil, M.~Kalaee, C.~P.~Dietrich, A.~S.~Clark, M.~J.~Holleran, A.~S.~Safavi-Naeini, and O.~Painter,
"Superconducting qubit to optical photon transduction,"
\href{https://doi.org/10.1038/s41586-020-3038-6}{\textit{Nature} \textbf{588}, 599--603 (2020).}

\bibitem{ref16}
T.~C.~van~Thiel, M.~J.~Weaver, F.~Berto, J.~M.~Fink, and S.~Hong,
"Optical readout of a superconducting qubit using a piezo-optomechanical transducer,"
\href{https://doi.org/10.1038/s41567-024-02742-3}{\textit{Nat. Phys.} \textbf{21}, 401--405 (2025).}

\bibitem{ref17}
R.~Nongthombam, P.~K.~Gupta, and A.~K.~Sarma, 
"Quantum transduction of a superconducting qubit in an electro-optomechanical and an electro-optomagnonical system,"
\href{https://doi.org/10.1103/PhysRevA.108.043501}{\textit{Phys. Rev. A} \textbf{108}, 043501 (2023). }.

\bibitem{ref18}
D.~Mishra and P.~Kumar, "Controlling nonadiabatic dynamics in an optomechanical array via the phase of the driving laser field,''  \href{https://doi.org/10.1364/JOSAB.572054}
{\textit{J. Opt. Soc. Am. B} \textbf{42}, 2296--2305 (2025).}

\end{thebibliography}
\end{document}